\documentclass[11pt]{article}
\usepackage{graphicx}
\begin{document}
\begin{center}

{\bf SPECTRAL VARIATIONS OF Ae-Be HERBIG STARS IN THE Mon R1
ASSOCIATION}\\
\vspace{8mm}
 \large
 L.A. PAVLOVA,L.N. KONDRATYEVA and R.R. VALIULLIN

 \vspace{4mm}
 \it{Fessenkov Astrophysical Institute, 050020, Almaty,
 Kazakhstan}

 \ email:lara@aphi.kz\\

    \vspace{4mm}

\end{center}
\vspace{4mm} \small \noindent We present the change in the
H$\alpha$\rm\, emission-line profile of the spectra of some Ae-Be
Herbig stars. In the spectrum of VY Mon H$\alpha$\rm\,may have one
of three profile types: P Cyg, PCygIII or single line in
accordance with the brightness variations of the star.   HD259431
now shows a double H$\alpha$\rm\, profile with the red component
stronger than the blue component, while in the earlier
observations the blue peak was higher than the red peak. Finally,
the last H$\alpha$\rm\, profile of LkH$\alpha$\rm\ 215 is very
similar to that  obtained by Finkenzeller et al.\\
\vspace{6ex} \noindent{KEY WORDS\,}\,\rm\mbox{Ae-Be Herbig
stars;spectral profiles of H$\alpha$\rm}

     \normalsize
\noindent One of the major problems caused by the Ae-Be Herbig
stars is their strong activity. There are emission lines of
different ionized elements in the spectra of  most of these stars.
The study of emission line profiles of Ae-Be Herbig stars is often
used to investigate the    physical conditions in the emitting
regions. H$\alpha$\rm\ emission remains  one of the best probes of
the inner circumstellar environments of young stars and contains
information about   the kinematics, excitation, and geometry of
the gas. The large infrared and submillimetre continuum excesses
and sometimes veiled photospheric absorption are indicators of the
presence of  cool circumstellar material with a complicated
structure \cite{He98}. The interpretation of observational data
critically depends on whether the circumstellar matter is
organized in a flat disk or in spherical envelopes \cite{CoH79,
Cas90,Bou98,Her04}. The detailed modelling of emission-line
profiles, mainly for the optically thick lines, such as
H$\alpha$\rm\, is very sensitive to the radiative transfer, level
populations, density and velocity field. However the evidence that
H$\alpha$\rm\, emission may be the result of the combined
contributions of more than one region is very important. The
envelope structure can be complicated by the presence of the
magnetic fields both of a  star and of the circumstellar
environment, and can be changed with time by different  parameters
on the line of sight . The analysis of the dusty environment is
used  to develop  an empirical model of the formation and
evolution of pre-main sequence stars. The main sources of
variability are unknown, and it is difficult to separate the
influence of stellar activity and circumstellar matter effects.
More complete understanding of the variety of line-forming regions
requires the study of the profiles of many lines of different
elements. Only a long-term set of observations for very many Ae-Be
Herbig stars can help us to understand this phenomenon.

The emission profiles of  H$\alpha$\rm\, can be classified as
double peaked, single or P Cygni. They show variability from a few
hours up to an year. Some stars are known to change the
H$\alpha$\rm\, profile from one type to another, but the moment of
such an event is unpredictable\cite{Fin84, Fer95, Rei96}. The
highly variable component of H$\alpha$ line profile affects mainly
the blue  and the central regions of  the profile and the time
variable must be connected with  the change in the condition in
the envelopes or with a mechanism of variability such as a
rotating star or envelope, mass loss, infall,  pulsation
\cite{ Bou98, Fi84}. The long-term  study  of the variations in the
emission profiles can provide a definition  of  the nature of the
stars.

 Our main goal is to  search  for the characteristic timescales and amplitudes
 of  the variability of  emission line details in order to
study  the arrangement of emission  regions and  the basic mechanisms of variations.
 We give now  the results of spectral observations on three stars in the MonR1 association:
 VY Mon, LkH$\alpha$\rm215 and HD 259431. Observations   were carried out with the 70 cm
 telescope of the Astrophysical Institute,
(Almaty, Kazakhstan) over  20 years. The apparatus  and  the
method of processing the observations have been  described in
\cite{Kon03}.

VY Mon is one of the  youngest  stars, with a large reddening
(A$_V $\rm\, of magnitude not less than 7.0) with a very high
degree of polarization ( about 10\%)\cite{Pav85} and with visual
and infrared excesses. Its spectral class was identified with
large uncertainty: O9 -- F5 \cite{CoH79,Cas90,Her04}. Our
observation programme  began in 1986; the first spectrograms with
a resolution of 2.7\r{A} showed an H$\alpha$\rm\, emission line
with a P Cyg peculiarity. In 1988 and 1991 a change in profile
type was displayed, from P Cyg  to P CygIII  ( because an
additional blue emission line had  appeared instead of  P Cyg
absorption), and then  another change, to a single-emission-line
profile,  was recorded \cite{Pav95}. When VY Mon is seen as a
bright star, the P Cyg peculiarity  is observed, and when the star
becomes weaker, the profile turns into P CygIII and then into a
single-line profile. The light curve of VY Mon presented in
\cite{Mir92} is quite similar to those of Herbig Ae-Be stars of
Algol type. However VY Mon shows a strong variation in color index
without significant changes in brightness . The velocity value of
P Cyg absorption varies from -62km s$^{-1}$ to -340km
s$^{-1}$\rm\,; the initial increase is replaced by a decrease over
8 month. We have derived the systematic values of the shift of the
main emission centre for different types of profile: -62km
s$^{-1}$ for P Cyg, -120km s$^{-1}$ for P CygIII, -140km s$^{-1}$
(for the single line).  However, the variations in the intensity of
the main emission relative to the continuum and to the red edge of
the line are weak.(figure 1). This can be connected with the
variation in  optical thickness without morphology changes. All
three types of profile imply that high-velocity outflows can be
important envelope components and can be explained by  the same
limited models of anisotropic stellar wind, with a variable
terminal velocity.

LkH$\alpha$\rm\, 215 is a star of spectral type B1-B7e with a very
broad double emission line H$\alpha$\rm. The emission lines CaII K,
CaII (8542 and 8498\r{A})  and OI (8446\r{A})  have identical type of
profiles \cite{Ham92}. In \cite{Fin84} for the separate components
of the H$\alpha$\rm\, profile the following values of radial
velocity were obtained:  -550 -120 - 35 +105 and +450km s$^{-1}$ -
for the blue edge, peak, central absorption, peak and red edge
respectively. Our estimations are somewhat different:
 -575 -
32 +58 +141 and +510km s$^{-1}$\rm\, respectively. (figure 2) In
the case of the double-peaked lines the relative intensities of the red
and blue peaks change considerable and alternately reach the
maximal intensity.  Such mechanisms as outflow (blue peak less) or
infall (red peak less) of envelope matter may be suggested as the
 interpretations of these changes.  The central absorption varies near zero point
with a small amplitude.  This could be connected  with the
variable thickness of matter on the line of sight. It is known
that the large majority of Herbg Ae-Be stars show deep central
absorptions, and this fact demonstrates
 that  equatorial circles  of obscuring material can be thick and close to the line of sight.

HD259431 is a star of  spectral type B6pe-AOe
 \cite {CoH79,Her04, Fi84, Ham92} with the
broad double emission lines of H$\alpha$\rm\,, H$\beta$\rm\,,
H$\gamma$\rm\, and so on.  The early observations   showed only
H$\alpha$\rm\,   with  a strengthened
 blue peak unlike our 2005 data which presented the opposite  picture; however  the radial
  velocity parameters of the
profile  are the same. The emission lines CaII K (3933\r{A}), CaII
(8542, 8498\r{A}),  OI (8446\r{A}) \cite{Ham92}  and MgI (2790\r{A})
have double-peaked profiles.  The identical  types of line profile
of different elements may  proves   that all of them are formed in
the same  region, near the star, inside an asymmetric disk -like
envelope. Usually  NI (8629\r{A}) line is detected in the hotter B
stars, because of its high ionization (14.5 eV) and
excitation(10eV) energy . The correlation between HeI, 4471 \r{A}
and MgII (4481\r{A})   confirms
 the B6e spectral class, as a favorable.
The spectral variations are very irregular(figure 3). A short time
variability is related to a region  closer to the star and
reflects a change in the  velocity field. Some stars are known to
change H$\alpha$\rm\, profiles from one type to another, but this
fact is not connected with their masses or  evolution status
\cite{Fin84, Fi84,Yud00}. All three stars can have disk-like
envelopes, however the types of profiles are different and do not
depend  on extinction or spectral class, but  do depend on the
thickness of the toroidal envelope and its inclination to the line
of sight.

\newpage
\begin{picture}(300,180)
\put(0,180){\special{em:graph fig.bmp}}
\end{picture}
\newpage

\end{document}